\documentclass[preprint]{aastex}

\shorttitle{Dickel,Goss,$&$ DePree}
\shortauthors{H$_2$CO absorption towards W~58~C1}

\begin{document}


\title{WSRT and VLA Observations of the 6~cm and 2~cm lines \\
of H$_2$CO in the direction of W~58~C1(ON~3) and W~58~C2}


\author{H{\'e}l{\`e}ne R. Dickel\altaffilmark{1,2}}
\affil{Astronomy Department, University of Illinois,,
1002 W. Green Street, Urbana, IL 61801}
\email{lanie@astro.uiuc.edu}

\author{W. M. Goss}
\affil{National Radio Astronomy Observatory
 P.O. Box O, Socorro, NM 87801}
\email {mgoss@nrao.edu}

\and

\author{C. G. De Pree}
\affil{Department of Physics and Astronomy, Agnes Scott College, 
141 East College Avenue, Decatur, GA 30030}
\email {cdepree@herschel.agnesscott.edu}


\altaffiltext{1}{Visiting Astronomer, ASTRON, 
P.O. Box 2, 7990AA Dwingeloo, The Netherlands}
\altaffiltext{2}{Visiting Professor, Astronomical Institute
'Anton Pannekoek', University of Amsterdam, Kruislaan 403, 1098SJ,
Amsterdam, The Netherlands}


\begin{abstract}
Absorption in the $J_{K_-,K_+}$ = $2_{11}-2_{12}$ transition of 
formaldehyde  at 2~cm towards the ultracompact HII regions
C1 and C2 of W~58 has been observed with the Very Large Array with an 
angular resolution of $\sim$ 0.2$''$ and a velocity resolution
of $\sim$1 km~s$^{-1}$. The high resolution continuum image of 
C1 (also known as ON~3) shows a partial shell which opens to the
NE. Strong H$_2$CO absorption is observed against W~58~C1.
The highest optical depth ($\tau$ $>$ 2) occurs in the SW portion of 
C1 near the edge of the shell, close to the continuum peak.
The absorption is weaker 
towards the nearby, more diffuse compact HII region C2, $\tau$~$\le$~0.3.
The H$_2$CO velocity ($-$21.2 km~s$^{-1}$) towards C1 is constant and
agrees with the velocity of CO emission, mainline OH masers,
and the H76$\alpha$ recombination line,
but differs from the velocity of the 1720 MHz OH maser emission 
($\sim$~$-$13~km~s$^{-1}$).

Observations of the absorption in the $J_{K_-,K_+}$ = $1_{10}-1_{11}$
transition of formaldehyde  at 6~cm towards
W~58 C1 and C2 carried out earlier 
with the Westerbork Aperture Synthesis Telescope  
at lower resolution ($\sim$4$''$x7$''$) show comparable optical depths
and velocities to those observed at 2~cm.
Based on the mean optical depth profiles at 6~cm and 2~cm,
the volume density of molecular hydrogen n(H$_2$) and the formaldehyde column 
density  N(H$_2$CO) were determined.
The n(H$_2$) is $\sim$6x10$^4$ cm$^{-3}$ towards C1.
N(H$_2$CO) for C1 is 
$\sim$8x10$^{14}$ cm$^{-2}$  while that towards C2 is 
$\sim$8x10$^{13}$ cm$^{-2}$.

\end{abstract}
  



\section{Introduction}
The star-forming complex \objectname[w58]{W~58} \citep{w58} at 
$\ell$ $\sim$70.3 and $b$ $\sim$ $+$1.6
is located on the far side of the Cygnus spiral arm at a distance of
8.7 Kpc \citep{i76}.
The region has been imaged in the continuum by \citet{w-w69}
and \citet{i76}, in CO by \citet{i80}, and in HI by \citet{r81}. 
The complex includes the optical nebulae NGC~6857 and 
\objectname[pk67]{K3-50} \citep{pk67}. At radio wavelengths,
\citet{w-w69} detected three main components
A, B, and C.  Component A coincides with optical nebula K3-50.
At high resolution ($<$4$''$), component C is resolved 
into ultracompact HII regions (UCHII regions) \objectname[h75]{W~58~C1} 
and \objectname[h75]{W~58~C2} \citep{h75}. 
Additional faint extensions to the emission from C1 are designated
C3 (to NW) and C4 (to SE) by \citet{cs77}. 
\citet{erw69} and \citet{rt69} independently discovered 
a 1720 MHz OH maser which was later 
designated \objectname[w70]{ON~3} \citep{w70}. 
\citet{w-www74} determined an accurate position for ON~3 
which can be associated with the UCHII region C1.

\citet{rsy79} observed the hydrogen Brackett $\alpha$ line at 4 $\mu$m 
towards W~58 and derived a visual extinction of $\sim$ 100 magnitudes 
towards C1 and $\sim$ 25 magnitudes towards C2. Van Gorkom et al. (1981)
observed W~58 in the H109 $\alpha$ recombination
line at 6 cm with the Westerbork Radio Synthesis Telescope (WSRT)
with an angular resolution of $\sim$ 10$''$.  The H109 $\alpha$ line towards C1
has an LSR velocity of $-$16 $\pm$ 3 km~s$^{-1}$, compared to a 
typical velocity of $-$22 km~s$^{-1}$ for the neutral gas [e.g. CO,
\citet{i80}]. From this observed velocity difference and the high extinction
seen towards component C1, both \citet{i80} and \citet{vsp81} favored the 
``blister'' model for the overall molecular cloud with UCHII
regions C1 and C2 located on the far side of the cloud. 
\citet{hpfd96} reached the same conclusion from near-infrared
observations
of the continuum emission and Brackett $\alpha$ and $\gamma$ lines.

Observations with the Very Large Array (VLA) at 2~cm with $\sim$
2$''$ resolution by \citet{dgp94}
reveal a core-halo structure in the continuum for C1 
and with a velocity of the H76 $\alpha$ line of $-$22 $\pm$ 2 km~s$^{-1}$,
in agreement with the velocity of the molecular gas. VLA observations
at an even higher resolution of $\sim$ 0.1$''$ by \citet{tm84}
show a shell-type structure in the continuum similar to that of W~3(OH),
but with a break in the shell to the east.   

\citet{fgd81} observed W~58~C in the $J_{K_-,K_+}$ = $1_{10}-1_{11}$ 
transition of formaldehye
at 6~cm with the WSRT with a resolution of $\sim$ 10$''$.  They
found strong H$_2$CO absorption towards C1 with an optical depth 
$\tau$ of $\sim$2 and weaker absorption with $\tau$ $\sim$0.25 towards C2.
W~58~C1 (ON3) exhibits the
deepest H$_2$CO absorption of any of the UCHII
regions observed by Forster at 6~cm with $\sim$5$''$ resolution 
\citep{fgd82, fb82}. In this paper, we present 
observations of the $J_{K_-,K_+}$ = $2_{11}-2_{12}$ transition of
H$_2$CO at 2~cm towards 
both C1 and C2 made with the VLA with a resolution of $\sim$0.2$''$.  These
observations were obtained in order to determine whether
the molecular gas towards C1 has the same toroidal structure as 
had been determined for the shell source W~3(OH) by \citet{dg87}.
The new 2~cm observations are described in \S\ 2 and the data are presented 
in \S\ 3.  In \S\ 4 mean H$_2$CO column densities and H$_2$ volume
densities in the molecular gas towards the UCHII regions are
derived from a comparison of the 2~cm and 6~cm H$_2$CO profiles,
using the same procedure as in previous papers of this series 
[e.g. \citet{dg87,dgc96}]. The nature of W~58~C1 is discussed in \S\ 5 and
the conclusions are summarized in \S\ 6. 

\section{Observations and Data Processing}

Observations of the ortho-formaldehyde absorption towards the UCHII 
regions C1 and C2 of W 58 were made in the $J_{K_-,K_+}$ =
$1_{10}-1_{11}$ (hereafter 6~cm H$_2$CO) and $J_{K_-,K_+}$ = $2_{11}-2_{12}$
(hereafter 2~cm H$_2$CO) transitions. The 6~cm transition was observed
with WSRT in 1980 by \citet{fgd82} with a resolution of
7.2$''$x3.8$''$ (PA=0$^\circ$).  
The calibration and 
processing of the WSRT data are discussed by \citet{fb82}.

The 2~cm VLA data were calibrated in the spatial frequency ($\it{u,v}$)
domain in the standard manner and then imaged using the 
``Pipeline'' system \citep{b81}. Uniform weighting was used during
the Fourier transformation resulting in a synthesized beam of 
0.11$''$x0.10$''$ (PA=$-$88.5$^\circ$).  
In addition, the $\it{u,v}$ data were  self-calibrated using the 
continuum image and then Fourier transformed 
with natural weighting resulting in a synthesized beam of 
0.25$''$x0.22$''$ (PA=$-$84.0$^\circ$). The instrumental parameters are
listed in Table 1 
for both the 6~cm and 2~cm data sets. The 2~cm data were processed and
analyzed using the Astronomical Image Processing System (AIPS) software
of NRAO.  No primary beam correction was applied because the field
of view is centered on C1 and the intensity of C2 is diminished by 
a factor of only a few percent. The coordinates for the field center
 are given in Table 1 for both J2000.0 and B1950.0 (as originally
specified for the observations).  A simple, yet accurate,
conversion from B1950.0 to J2000.0 was applied to the small 
regions around C1 and C2.  Only the larger image
of the entire region (Fig.~1a) remains with B1950.0 coordinates;
this provides an easy comparison with earlier images.

\placetable{table1}

Two ``CLEANED'' continuum images were made at 2 cm: one with uniform weighting
to obtain the highest resolution and the self-calibrated image with 
natural weighting to obtain the best sensitivity to the
 more extended
emission. Channel images of the absorbed continuum 
due to the H$_2$CO molecule were made for the 2 cm line.

Images of the H$_2$CO optical depths ($\tau$) were obtained from 
the continuum image ($T_c$) and from the channel images of the absorbed
continuum at various velocities ($T_{\ell}$):  
$\tau$ = -ln($T_{\ell}$/$T_c$). Optical depths were calculated 
only where the signal-to-noise ($S/N = \tau$/$\sigma_{\tau}$) was $>$ 2.
Because of the weakness of the continuum,  the sensitivity of the line 
data at the original resolution of $\sim$0.1$''$ was inadequate.
Therefore, optical depth images with a resolution of 
0.24$''$ were made.

The 2 cm VLA data are available on the Web at \\
http://adil.ncsa.uiuc.edu/document/00.HD.01

\section{Data}

\subsection{Continuum Emission}

Continuum images at 2 cm of W~58~C1 (to the west) and C2 (to the east) 
made with natural weighting
are shown in Fig.~1a.  The rms $\sigma$ is 0.5 mJy beam$^{-1}$ (beam
$=$ 0.25$''$x0.22$''$). Fig.~1a shows  
 low-level extended emission surrounding the compact core of C1 which
corresponds to the C3 (in NW) and C4 (in SE) components of \citet{cs77}.
Within the 3 $\sigma$ level, the integrated flux density is 
400~$\pm$~30~mJy for C1 and 190~$\pm$~40~mJy for C2. These flux
densities are somewhat lower than the values of 500~$\pm$~10~mJy for C1 and 
200~$\pm$~40~mJy found by \citet{dgp94} at the lower resolution  
of 1.4$''$x1.1$''$. The VLA in A array at 2~cm is insensitive to 
structures larger than about 4$''$.  If the missing flux density of
$\sim$100~mJy for C1 were spread over a 4$''$ region, then the additional
intensity would be equivalent to $\sim$0.7~$\sigma$ and thus not
detected.  

\placefigure{fig1}

A new small source, designated C5 in Fig.~1a, is located to 
the south of C1 and C2 at (B1950) $\alpha$ = 19$^h$59$^m$58.74$^s$
$\pm$0.01$^s$ and $\delta$ = 33$^{\circ}$25$'$36.2$''$
$\pm$0.1$''$ [(J2000) $\alpha$ = 20$^h$01$^m$54.34$^s$
and $\delta$ = 33$^{\circ}$34$'$00.8$''$].  
This source has a peak flux density of 
3.5$\pm$0.5 mJ~beam$^{-1}$
and an integrated flux density of 18$\pm$2.8 mJy.
It's size is $\sim$0.7$\pm$0.1$ ''$x0.4$\pm$0.05$''$  
(PA$\sim$110$^\circ\pm$9$^\circ$). \citet{bgd00} measured a flux density
of 20 mJy at 3.6~cm for this source
which gives a spectral index $\alpha \sim -$0.1.

A continuum image of UCHII region C2 is shown as both grey-scale
and with contours in Fig.~1b with
a resolution of 0.22$''$.  C2 appears as a ridge of emission
 elongated almost north-south 
with more diffuse emission to the west.  

The highest resolution image (0.11$''$) of UCHII region C1 in Fig.~1c
clearly shows the partial
shell structure with the break to the east. With a  sensitivity 
of~1 mJy beam$^{-1}$ the extended emission is not detected.
Contours of the extended emission in the natural weighted image
(beam $=$ 0.25$''$x0.22$''$) are shown in Fig.~1d, superposed on a
grey-scale presentation of the uniform-weighted, higher-resolution image.

\subsection{Formaldehyde Absorption at 2 cm}

\subsubsection{Spatial Distribution}

The 2~cm H$_2$CO optical depths towards the UCHII region W~58~C1 with a
resolution of 0.24$''$ are presented for nine velocities in Fig.~2
as grey scale images with contours superposed. The S/N ranges from
about two at the edges to six~-~seven just east of the continuum peak. 
 The highest optical depths with 
$\tau(H_2CO) \ge$ 2 occur at a velocity $V_{LSR} = -$21.2
km~s$^{-1}$ along a north-south ridge running across and slightly west of the
continuum peak. 

\placefigure{fig2}
\placefigure{fig3}

The correspondence between H$_2$CO optical depth and 
continuum emission is illustrated in Fig.~3 where the 0.11$''$ 
resolution continuum contours for C1 from Fig.~1c are superposed on
the grey scale image of the optical depth for $V_{LSR} = -$21.2
km~s$^{-1}$. The close spatial correspondence between the high
formaldehyde opacity and the C1 continuum peak indicates that this
molecular gas is probably physically associated with the UCHII region.

\subsubsection{Velocity Structure}

The width of the H$_2$CO line at 2 cm  $\Delta$$V(FWHM)$ 
is between 2 and 3 km~s$^{-1}$ towards C1.
The velocity of the line is essentially
constant across the UCHII region with  $V_{LSR} \sim -$21.2
km~s$^{-1}$; there is a variation of, at most, $\sim$0.3 km~s$^{-1}$. 
At the position 
where the optical depth is highest, along a ridge just to the west of
the continuum maximum, the H$_2$CO velocity is slightly more
positive than the average ($<V_{LSR}>~= -$21.1 km~s$^{-1}$).  To the east
of the continuum maximum where there is a hole and break in the HII shell,
the H$_2$CO velocity is more negative ($<V_{LSR}>~= -$21.4
km~s$^{-1}$); here the molecular gas is expanding away from the 
UCHII region.

\subsection{Spatially integrated H$_2$CO Profiles at 6 cm and 2 cm}

Spatially-integrated H$_2$CO absorption profiles at 6 cm towards the 
continuum peaks
of W~58~C1 and C2 shown in Fig.~4 were taken from 
\citet{fgd82}. 
In order to compare the 6 cm and 2 cm data, the 2 cm H$_2$CO line data
were also integrated over each source - over the area encompassed
by the outer contour of Fig.~1d for C1 and by the outer contour
of Fig. 1b for C2.
These spatially integrated profiles are shown in Fig.~5. 

\placefigure{fig4}
\placefigure{fig5}

\section{Mean H$_2$ Densities and H$_2$CO Column Densities}

\subsection{Determination of $n(H_2)$ and $N(H_{2}CO)$}

The method described by \citet{dgrb86} and \citet{dg87}
was used to obtain the H$_2$ spatial density $n(H_2)$, and the H$_2$CO
column density $N(H_{2}CO)$.  The procedure is based on 
radiative transfer calculations using the Sobolev 
approximation. In the H$_2$ density range
10$^3$ $\le$ $n(H_2)$ $\le$ 10$^6$, 
the strength of the H$_2$CO absorption at 2~cm relative to that
at 6 cm is related to the H$_2$ density
while the strength of the absorption at 6~cm is sensitive to 
variations in H$_2$CO column density.

Input parameters for the radiative transfer calculations 
of \citet{dgrb86} are
the  kinetic temperature $T_K$, the 
molecular hydrogen density $n(H_2)$,
the radius of the cloud $R$,  maximum velocity $V$, and
the number density of ortho H$_2$CO molecules in all levels $n_t$.
After the populations of the levels $x_j$ were determined for
particular values of $T_K$ and $n(H_2)$ and a range of $n_t$, the
expected optical depths $\tau_{jk}$ for the 6~cm and 2~cm lines were
calculated according to the following relationship:
\begin{equation}
\tau_{jk} = [\frac{c^3}{8\pi}\frac{A_{jk}g_j}{\nu^3_{jk}}] \cdot [x_k -
x_j] \cdot [\frac{n_t}{(dv/dr)}]
\end{equation}
or
\begin{equation}
\tau_{jk} \propto [x_k - x_j] \cdot n(H_2) \cdot A(orth H_2CO),
\end{equation}
where the relative abundance per velocity gradient $A(orth~H_2CO)$ is
given by
\begin{equation}
A(ortho~H_2CO) = \frac{[ortho~H_2CO]}{[H_2]} \cdot \frac{1}{(V/R)} = 
\frac{n_t}{n(H_2)} \cdot \frac{1}{(dv/dr)}.
\end{equation}
\citet{dg87} present curves of constant $n(H_2)$ and constant $A(ortho~
H_2CO)$
as functions of $r=\frac{\tau(6cm)}{\tau(2cm)}$ and $\tau(6cm)$
for $T_K$ = 40 K and $T_K$ = 60 K.

With [H$_2$CO]/[ortho~H$_2$CO] $=$ 1.33 and $A$ in units of 
(km s$^{-1}$ pc$^{-1}$)$^{-1}$,~ 
$N_V(H_{2}CO)$ in units of cm$^{-2}$ (km s$^{-1}$)$^{-1}$ is defined as
\begin{equation} 
N_V(H_{2}CO)  = 4.1~10^{18} \cdot A(ortho~H_2CO) \cdot n(H_2).
\end{equation}
Multiplying $N_V(H_{2}CO)$ by the FWHM line width $\Delta V$, 
yields the H$_{2}$CO column density, $N(H_{2}CO)$.

There is no measurement of the kinetic temperature towards C1 but
a radiation temperature of 26~K (with $T_K$ $\sim$ 30 K)
has been observed by \citet{pm91} in the 
CO J=$3-2$ transition with 30$''$ resolution toward K3-50 ($\sim$2$'$ to SW of
C1).  Other molecular clouds with embedded HII regions
such as DR~21, W~49~A~north, and W~51~A have 
kinetic temperatures between 40~K and 50~K \citep{pwr88}.
Fortunately, the radiative transfer calculations are insensitive 
to kinetic temperature. 
For example, changing $T_K$ by 20 K changes 
$n(H_2)$ and $N(H_2CO)$ by less than 15$\%$ whereas
observational uncertainties of 10$\%$ in $\tau_{jk}$ can   
produce uncertainties on the order of 50$\%$ in these 
quantities.  Therefore, the radiative transfer results
for $T_K =$ 40 K from \citet{dg87} were used to determine
$n(H_2)$ and $N_V(H_{2}CO)$ for C1.

\subsection{Results for W~58~C1 and C2}

It is not possible to determine values of $n(H_2)$ and $N_V(H_{2}CO)$
for different regions of the sources because of the limited resolution 
of the 6 cm observations.  Therefore, mean values were derived
for W~58~C1 and C2 from the spatially integrated profiles shown in
Figs.~4 and 5.  The resulting $n(H_2)$ and $N_V(H_{2}CO)$ for C1
are shown in Fig.~6.  Although the sensitivity is insufficient to obtain 
a good determination of $n(H_2)$ for C2, an estimate of $N_V(H_{2}CO)$ 
can be determined from the 6 cm optical depth because of its dependence 
on H$_{2}$CO column density.

\placefigure{fig6}

The mean H2 density for the molecular gas in front of C1 is in the range
(4$-$8)x10$^4$~cm$^{-3}$ and the H$2$CO column density is  
$N(H_{2}CO) =$ (8$\pm$4) x10$^{14}$ cm$^{-2}$. 
The corresponding value of the column density of the
molecular gas in front of C2 is $\sim 10^{14}$ cm$^{-2}$.

Using the \citet{js74} relationship between visual
extinction $A_V$ and $N(H_2)$, values of $N(H_2)$ were derived 
from the $A_V$ given by \citet{rsy79} ($A_V \sim$ 25 magnitudes 
in front of C2
and $\sim$ 100 magnitudes in front of C1).
The derived relative abundance [H$_2$CO]/[H$_2$] of about 
2x10$^{-9}$ in the direction of C2 is similar to the (1$-$2)x10$^{-9}$
determined for W~3A, W~49A north and south, and DR~21A
whereas the higher relative abundance of $\sim$6x10$^{-9}$ towards C1
is closer to the $\sim$5x10$^{-9}$ found for the torus around W~3(OH)
and towards DR~21C [see \citet{dg90}].

\section{The Nature of W~58~C1}

Following \citet{wc89}'s classification scheme for  UCHII regions,
W~58~C1 is either a cometary UCHII region like W~49~S \citep{gm87} or a broken
shell UCHII region such as W~3(OH) \citep{dg87}. These two scenarios
are explored in the next two sections (\S\S\ 5.1. and 5.2.). 

\subsection{Cometary UCHII Region - Comparison with Bow-Shock Models}

According to models explaining cometary UCHII regions as stellar-wind 
bow shocks, the ionizing star of C1 is expected to be
moving towards the southwest and observations at higher resolution 
should show a steep gradient in the continuum emission on the southwest side
 [\citet{vmwc90}; \citet{vm92}].  
In addition, there should be a relative 
velocity of approach between the molecular and ionized gas of 
several km~s$^{-1}$ depending on the velocity of the ionizing star.
From spatially integrated H76$\alpha$ radio recombination line
profiles, \citet{dgp94} find a mean velocity of the ionized gas of
 $V_{LSR} = -22$ $\pm$ 2 km~s$^{-1}$, 
comparable to that of the molecular gas~~-~
indicating no relative motions between them.

The OH masers are expected to be located in the shocked shell 
(not in the pre-shock material).  In this case, the measured velocity
of the 1720 MHz OH masers [$V_{LSR} \sim -$13~km~s$^{-1}$ \citep{w70}]
would indicate that the star is moving not only to the west but also away from 
the observer. Because the  accuracy of the 1720 MHz OH position is 
insufficient to precisely locate the masers relative to the continuum source,
no firm conclusions can be made at this time regarding the 
applicability of the stellar-wind, bow shock model to C1.

\subsection{Broken Shell - Comparison with W~3(OH)}

With the current data, the most favorable  model for C1 is that of a 
broken shell, similar to the shell source W~3(OH). Both the 6~cm and the 2~cm
H$_2$CO absorption lines towards W~3(OH) were imaged using the VLA
by \citet{dg87}.  As in the case of W~58~C1, 
pronounced changes in optical depth across W~3(OH) are also observed.
 
For W~3(OH) ~-~ at the velocities of the OH masers, the 2~cm H$_2$CO absorption
only covers the western half of the W~3(OH) in contrast to
the more ubiquitious 6 cm H$_2$CO absorption.  The data
are consistent with a toroidal model for the dense molecular
gas surrounding W~3(OH). The shell appearance of the HII region
is due to the ionization of the molecular torus.
The H$_2$ density is much higher in 
the torus than in the surrounding envelope of molecular gas.
An accurate determination of the H$_2$ density in the torus is not possible
from the mean optical depths alone; this requires detailed imaging in two 
transitions [as was done by \citet{dg87} for W~3(OH).]  

For W~58~C1 ~-~ the optical depth of the 2~cm H$_2$CO absorption varies across
this source with the highest values associated with a ridge at the 
southwest edge of the bright continuum arc.
Observations towards the region of the SE break in the continuum shell
indicate that the molecular gas may be expanding  away 
from the UCHII region towards the observer.
The derived mean H$_2$ density of the absorbing gas towards C1 
is similar to that found in the overall
molecular cloud surrounding W~3(OH), while the mean relative H$_2$CO 
abundance towards C1 is similar to that derived for the W~3(OH) 
molecular torus.

A significant increase in the H$_2$ density
where the 2~cm transition has its maximum optical depth would  
be further evidence that a molecular torus also surrounds C1.
High sensitivity and resolution spatial imaging in the 
1~cm transition of H$_2$CO in combination with the
existing 2~cm data is required for a determination of the H$_2$ density 
across C1.

The observations of the continuum emission and H92$\alpha$
observations at 3.6 cm  of \citet{bgd00} not only support the 
broken shell model for W~58~C1 but show evidence for a bipolar
outflow. Our observations of the continuum at 2~cm with
$\sim$0.2$''$ resolution (Fig.~1a) reveal an 
elongated envelope in the NW to SE direction, corresponding
to components C3 and C4 of \citet{cs77}.  The 3.6 cm data
with 2$''$ resolution show that the ionized gas extends to 
more than twice the distance shown in Fig.~1a. There is a velocity
gradient in the H92$\alpha$ recombination lines 
from $V_{LSR} \sim -$18  km~s$^{-1}$ in the NW to  $\sim -$24
km~s$^{-1}$ in the SE indicating that this is a bipolar outflow
and that the systemic velocity of C1 is the same as that
of the molecular cloud ($V_{LSR} \sim -$21  km~s$^{-1}$).
Some H$_2$CO molecules are apparently entrained in the ionized outflow,
albeit at a reduced velocity; this expanding gas arises from the gap
in the UCHII shell.

\section{Conclusions}

H$_2$CO absorption has been observed at 2~cm with the VLA towards the 
UCHII regions C1 and C2 of the W~58 star-forming region.  Weak
absorption with $\tau$ $\le$ 0.3 is seen towards C2.
A prominent line with $\tau$ $>$ 2 is observed in the SW part
of C1 close to the continuum peak and is physically associated with 
this UCHII region.
The H$_2$CO line at 2~cm is narrow with $\Delta V\sim$~2~km~s$^{-1}$
and $V_{LSR}$~of $\sim-$21.2 km~s$^{-1}$.
This velocity is in good agreement with the 
H76$\alpha$ recombination line velocity ($-22\pm2$~km~s$^{-1}$).
The molecular gas in the east towards the gap in the UCHII shell 
is expanding by 
$\sim$ 0.3~km~s$^{-1}$ relative to the molecular gas with large
integrated opacity
near the western edge of the shell.

A value of $n(H_2)$ $\sim$6x10$^4$ cm$^{-3}$ and of
$N(H_2CO)$ $\sim$8x10$^{14}$ cm$^{-2}$ is determined for the absorbing
molecular gas in front of the UCHII region C1 from a comparison
of the mean H$_2$CO profiles at 6~cm and at
2~cm.  The $N(H_2CO)$ towards C2 is estimated to be 
0.1 that found towards C1.  The relative abundance
[H$_2$CO]/[H$_2$] is $\sim$6x10$^{-9}$ towards C1 and $\sim$2x10$^{-9}$
towards C2.

W~58~C1 exhibits a core-halo structure with the core appearing 
as a $''$broken shell$''$.  The 2 cm absorption is strongest 
at the edge of the bright arc of continuum emission and suggests
that the molecular gas might also be in a dense toroid around the 
UCHII similar to W3(OH).  To confirm this
will require sensitive, high resolution imaging of the 1 cm transition 
of H$_2$CO.  
To completely rule out the possiblity of a $''$cometary$''$ 
(bow-shock) model
for W~58~C1, higher resolution imaging of 
radio recombination lines, the 1720 OH maser emission, and 
the continuum emission are required.

\acknowledgments

Partial support of the initial stages of this research from 
NATO grant 138.80 to HRD and WMG is gratefully acknowledged. 
HRD acknowledges partial support from the Laboratory of Astronomical
Imaging which is operated with funds provided by the Berkeley-
Illinois-Maryland Association; such research was partially supported 
by the National Science Foundation through grant 96-13999 to the 
University of Illinois. The National Radio Astronomy Observatory is 
a facility of the National Science Foundation operated under
cooperative agreement by Associated Universities, Inc..
The Westerbork Synthesis Radio Telescope is operated by the ASTRON
(Netherlands Foundation for Research in Astronomy) with support
from the Netherlands Foundation for Scientific Research NWO.
HRD wishes to thank the Astronomical Institute of the University 
of Amsterdam and the ASTRON for their hospitality during the final 
stages of this project. CGD thanks NRAO for hospitality in 
Socorro, NM where some of this work was carried out.

\clearpage



\figcaption[hdickel.fig1a.eps,hdickel.fig1b.eps,hdickel.fig1c.eps,
hdickel.fig1d.eps]
{Continuum images of the W~58~C region made with the VLA at a wavelength of
  2~cm.  Positive contours are solid lines and negative ones are dashed.
  The first figure (1a) is with B1950.0 coordinates; 
  all subsequent figures are with J2000.0 coordinates.\\
a) Natural-weight, self-calibrated continuum image of W~58~C at a
  resolution of 0.25$''$x0.22$''$ (PA=$-84.0^\circ$). 
  UCHII region C1 and C2 are to the west and east, respectively.
  Contour levels are -4, -2, 2 ($\sim4\sigma$), 4, 8, 16, 32,
  and 64 mJy beam$^{-1}$. The (natural-weighted) beam is shown in the
  lower left corner.\\
b) Grey-scale image plus contours of the emission from UCHII region
  W~58~C2 with a resolution of 0.22$''$. Contours are -2.5, 2.5
  ($\sim3\sigma$), 4, 5.5, and 7 mJy beam$^{-1}$. The
  (uniform-weighted) beam is shown at the lower left. \\
c) Continuum image of UCHII region W~58~C1 (ON~3) 
  with a resolution of 0.11$''$.  Contours
  are -2.5, 2.5 ($\sim3\sigma$), 5, 7.5, 10, 12.5, and 13.6 mJy
  beam$^{-1}$. The (uniform-weighted) beam is show at the lower left. \\
d) Overlay of three outer contours of the
  lower resolution (0.25$''$x0.22$''$) continuum image on the higher
  resolution (0.11$''$), grey scale image of the UCHII region
  W~58~C1. The contours are 5, 10, and 20 mJy beam$^{-1}$. The first
  contour is at the $\sim10\sigma$ level and encompasses the area over
  which the mean profile shown in Fig. 5 was obtained.
\label{fig1}}

\figcaption[hdickel.fig2.eps]{Selected channel images of the H$_2$CO optical
  depth in the $2_{11}-2_{12}$ transition at 2~cm towards UCHII
  region W~58~C1 shown in grey scale with
  contours superposed. The 0.24$''$ beam is shown in the first panel.
  The LSR velocity for the channel is given at the top of each
  panel. For the fourth panel with $V(LSR) =  -$21.2 km~s$^{-1}$, the
  contours are $\tau =$ 1.0 to 2.5 by 0.5, and
  2.9.  For panels 2, 3, and 5, the contours are $\tau =$ 0.5 and
  1.0. For the remainder, the contour is $\tau =$ 0.25.
\label{fig2}}

\figcaption[hdickel.fig3.eps]{Grey scale image of the 2~cm H$_2$CO optical
  depth at $V(LSR) = -$21.2 km~s$^{-1}$ towards W~58~C1 at a resolution
  of 0.24$''$ with continuum contours from Fig. 1c superposed.
  The optical depth scale is given at the top.
\label{fig3}}
\figcaption[hrd4.eps]{Spatially integrated profiles of the 6~cm
  H$_2$CO absorption towards W~58~C observed with the WSRT:
  towards component C1 at the bottom  and towards C2 at the top. 
  The left-hand side gives the flux density scale in mJy
  and the right-hand side gives the H$_2$CO optical depth scale.  
  The ($\pm~1~\sigma$) rms error bars are shown; 
  $\sigma = $11.3 mJy for both C1 and C2.
\label{fig4}}

\figcaption[hdickel.fig5.eps]{Spatially integrated profiles of the 2~cm 
  H$_2$CO absorption towards W~58~C observed with the VLA: 
  towards component C1 at the bottom and towards C2 at the top. 
  The left-hand side gives the flux density scale in mJy
  and the right-hand side gives the H$_2$CO optical depth scale. 
  The ($\pm~1~\sigma$) rms error bars are shown; $\sigma =$ 8.4 mJy for
  C1 and $\sigma =$ 11.5 mJy for C2.
 \label{fig5}}

\figcaption[hdickel.fig6.eps]{Profiles of mean H$_2$ volume density,
  $n(H_2)$, shown at the bottom 
  and mean H$_2$CO column density per velocity interval, $N_V(H_{2}CO)$, 
  shown at the top as a function of velocity towards W~58~C1. 
  The error bars are $\pm~1~\sigma$.  The dashed
  horizontal line in the upper panel indicates an approximate value
  for $N_V(H_{2}CO)$ for C2.  
\label{fig6}}





\begin{deluxetable}{lll} 
\tabletypesize{\footnotesize}
\tablecolumns{3} 
\tablewidth{0pc} 
\tablecaption{Instrumental Parameters \label{Table1} } 
\tablehead{ 
\colhead{}    &  \multicolumn{2}{l}{$\phm{WSRT}$Instrument}\\
\multicolumn{1}{l} {Parameter} & 
\multicolumn{1}{l} {WSRT} & 
\multicolumn{1}{l} {VLA}} 
\startdata 
Configuration & 3 km & A array \\
H$_2$CO transition & $1_{10}-1_{11}$ & $2_{11}-2_{12}$ \\
Rest frequency(MHz) & 4829.6596 & 14488.4788 \\
Wavelength (cm) & 6.21 & 2.07 \\
Primary Beam (') & 10.8 & 3.0 \\
Source & W~58~(entire) & W~58~C \\
Field Center & ~ & ~ \\
\phn \phn \phn \phn $\alpha$(B1950) & 19$^h$59$^m$54.00$^s$ & 19$^h$59$^m$58.48$^s$ \\
\phn \phn \phn \phn $\delta$(B1950) & 33$^{\circ}$24$'$30.0$''$
 & 33$^{\circ}$25$'$49.6$''$ \\
\phn \phn \phn \phn $\alpha$(J2000) & 20$^h$01$^m$49.61$^s$ & 20$^h$01$^m$54.07$^s$ \\
\phn \phn \phn \phn $\delta$(J2000) & 33$^{\circ}$32$'$54.24$''$ 
 & 33$^{\circ}$34$'$14.17$''$ \\
Dates of observations & 1980 October 19 & 1985 January 18 \\
Integration time (hr) & 12 & 9 \\
Number of telescopes & 14 & 25 \\
Number of channels & 31 & 31 \\
Velocities (LSR, km~s$^{-1}$) & ~ & ~ \\
\phn \phn \phn \phn Central vel (ch 16) & -21.4 & -21.2 \\
\phn \phn \phn \phn Resolution & 0.727 & 1.01 \\
\phn \phn \phn \phn Channel spacing & 0.606 & 1.01 \\
Synthesized beam ($''$) & 7.2x3.8 (PA$=0.0^{\circ}$) & ~ \\
\phn \phn \phn \phn Natural weight & ~ & 0.253x0.224 (PA$=-84.0^{\circ}$) \\
\phn \phn \phn \phn Uniform weight & ~ & 0.106x0.098 (PA$=-88.5^{\circ}$) \\

\enddata 
\label{table1}
\end{deluxetable} 
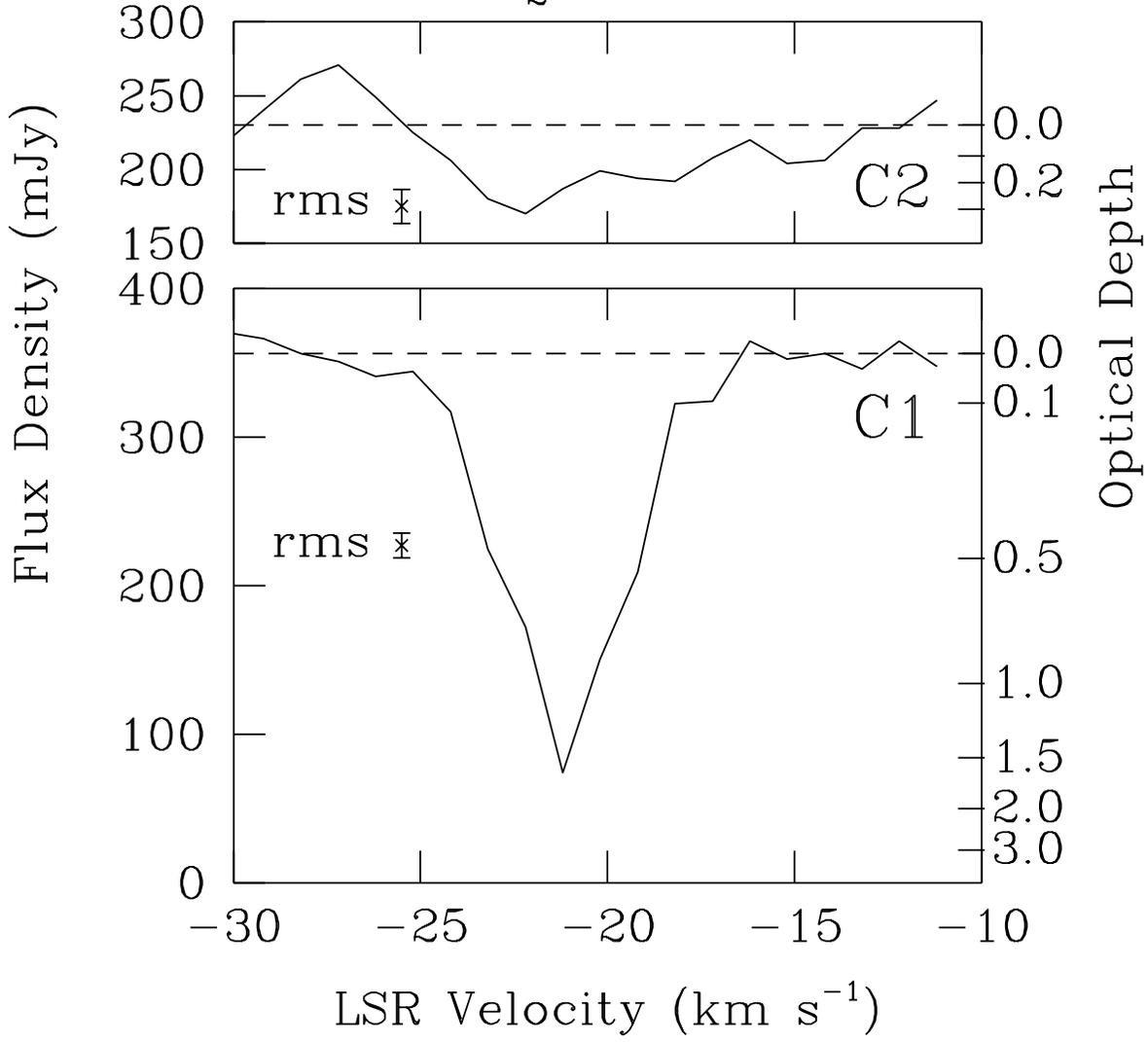
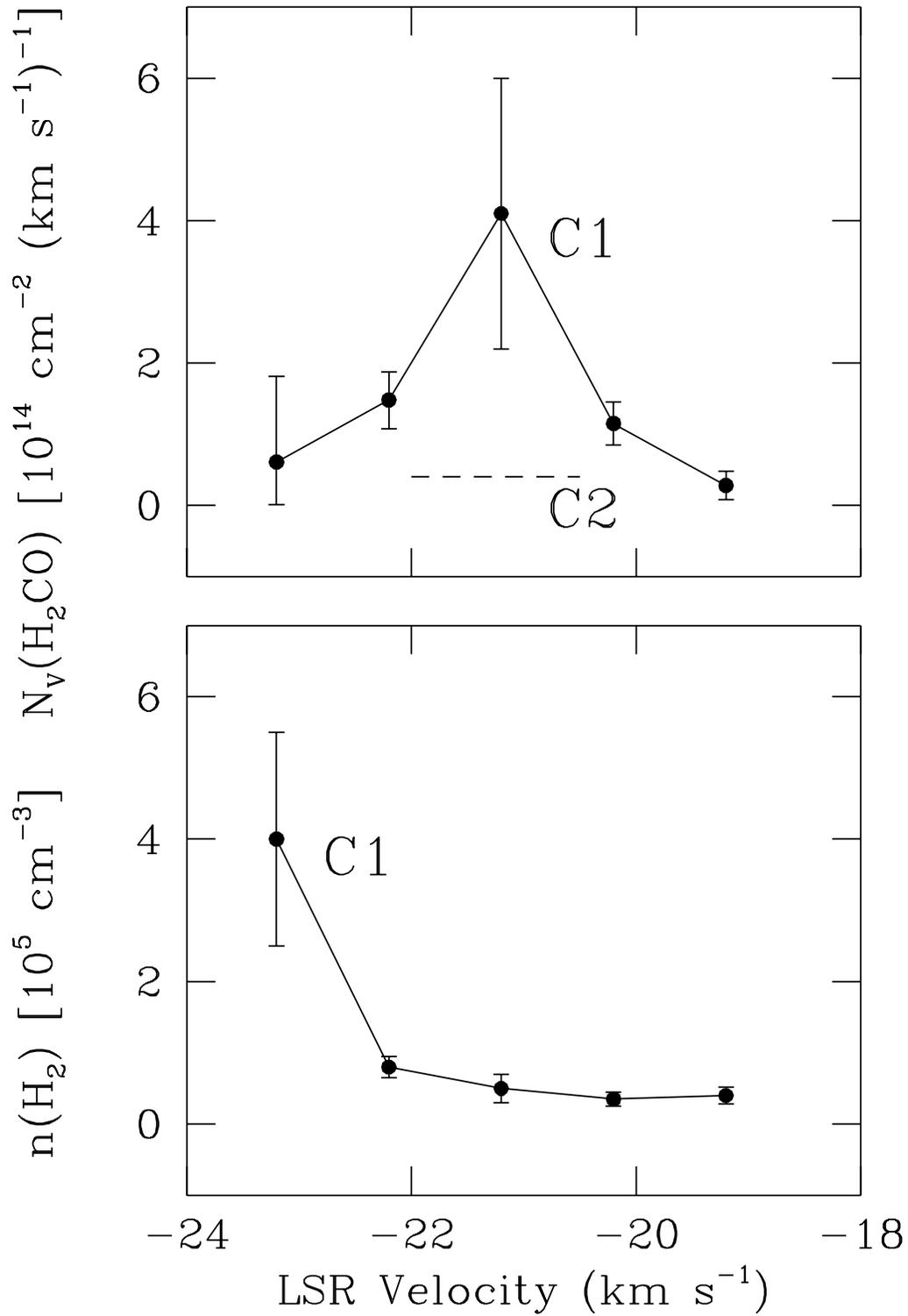
\end{document}